\documentclass[10pt,conference]{IEEEtran}

\usepackage{amsmath,amssymb,amsfonts}
\usepackage[]{algpseudocode}                               % algorithm package
\usepackage{algorithm}
\usepackage{float}

\renewcommand{\algorithmicrequire}{\textbf{Input:} }       % Use Input in the format of Algorithm
\renewcommand{\algorithmicensure} {\textbf{Output:}}       % Use Output in the format of Algorithm
\usepackage{graphicx}
\usepackage{textcomp}
\usepackage{xcolor}
\usepackage{circuitikz}
\usetikzlibrary{shapes}
\usetikzlibrary{positioning}

\usepackage{comment}
\ctikzset{logic ports=ieee}
\def\BibTeX{{\rm B\kern-.05em{\sc i\kern-.025em b}\kern-.08em
    T\kern-.1667em\lower.7ex\hbox{E}\kern-.125emX}}

\usepackage{biblatex}
\usepackage{fancyhdr}
\usepackage{lastpage}

\newif\ifpagenum
\pagenumfalse
\ifpagenum
\fi

\newif\ifrev
\revfalse

\ifrev
  \newcommand{\yier}[1]{{\color{green} [Yier: #1]}}
  \newcommand{\travis}[1]{{\color{blue} [Travis: #1]}}
  \newcommand{\jim}[1]{{\color{magenta} [Jim: #1]}}
  \newcommand{\shaojie}[1]{{\color{Fuchsia} [Shaojie: #1]}}
\else
  \newcommand{\directions}[1]{}
  \newcommand{\travis}[1]{}
  \newcommand{\yier}[1]{}
  \newcommand{\jim}[1]{}
  \newcommand{\shaojie}[1]{}
\fi

\newif\ifaddressed
\addressedtrue
\ifaddressed
  \newcommand{\addressed}[1]{}
\else
  \newcommand{\addressed}[1]{\textit{#1}}
\fi

\newif\ifcut
\cuttrue
\ifcut
  \newcommand{\cut}[1]{}
\else
  \newcommand{\cut}[1]{#1}
 \fi

\title{Improving FSM State Enumeration Performance for Hardware Security with RECUT and REFSM-SAT}

\author{
    James Geist$^{*}$, Travis Meade$^{*}$, Shaojie Zhang$^{*}$, Yier Jin$^{\dag}$\\
    \textit{$^{*}$Department of Computer Science, University of Central Florida}\\
    \textit{$^{\dag}$Department of Electrical and Computer Engineering, University of  Florida}\\ 
    james.geist@ucf.edu, travis.meade@ucf.edu, shaojie.zhang@ucf.edu, yier.jin@ieee.org\\ 
}

\begin{document}
\maketitle

\begin{abstract}
Finite state machines (FSM’s) are implemented with sequential circuits and are used to orchestrate the operation of hardware designs. Sequential obfuscation schemes aimed at preventing IP theft often operate by augmenting a design's FSM post-synthesis. Many such schemes are based on the ability to recover the FSM's topology from the synthesized design. In this paper, we present two tools which can improve the performance of topology extraction: RECUT, which extracts the FSM implementation from a netlist, and REFSM-SAT, which solves topology enumeration as a series of SAT problems. In some cases, these tools can improve performance significantly over current methods, attaining up to a 99\% decrease in runtime.

\end{abstract}

%
% 100 word version for submissions
%
% Finite state machines (FSM’s) are sequential circuits which orchestrate the operation of hardware designs. Sequential obfuscation schemes aimed at preventing IP theft often operate by augmenting a design's FSM post-synthesis. Many such schemes require the ability to recover the FSM's topology from the synthesized design. In this paper, we present two tools which can improve the performance of topology extraction: RECUT, which extracts the FSM implementation from a netlist, and REFSM-SAT, which solves topology enumeration as a series of SAT problems. In some cases, these tools improve performance significantly (up to a 99% decrease in runtime) over current methods.
%

\section{Introduction}

% Consumption of Integrated Circuits (ICs) dreamt by Natural Language Processing (NLP) models will become an unremarkable aspect of future System-on-Chips (SoCs). Cars, drivers, and driverless cars will continue to put a strain on IC production. ``Solutions'' that deter Intellectual Property (IP) pirates may be significantly weaker than previously thought.

% ChatGPT and other Natural Language Processing (NLP) models will significantly improve in the coming years, increasing the demand for hardware and the basic requirements to run such software.  Additionally, hardware applications and designs developed by NLPs may become commonplace among the fabricated Integrated Circuits (ICs)~\cite{nair2023generating}. Using AI generated chips will drastically increase the production of ICs. 

The field of Natural Language Processing (NLP) models is poised for significant improvements, resulting in a substantial rise in hardware demand. NLP improvements necessitate enhanced hardware production capabilities to accommodate the escalating computational requirements of these state-of-the-art software systems. Furthermore, NLP-designed hardware applications have been tested by researchers and are anticipated to enter the realm of fabricated Integrated Circuits (ICs)~\cite{nair2023generating}. The integration of AI-generated chips in IC production will substantially increase IC output, addressing an escalating demands for ICs.

The United States' 2022 CHIPS Act injected an estimated \$50 billion into IC research and manufacturing, boosting production to meet consumer product demands (e.g. cars, AR/VR devices). The influx of new devices, particularly those developed by NLPs, necessitates meticulous verification to ensure proper functionality, while increased IC consumption exacerbates revenue loss from Intellectual Property (IP) theft. IP theft, including chip overproduction, was projected to cost companies nearly a trillion dollars in 2017~\cite{economics2017economic}.

IC obfuscation can be used to help prevent IP theft due to overproduction. Both combinational~\cite{sengupta2022new, roshanisefat2018srclock, kamali2019full} and sequential~\cite{hu2020sanscrypt} methods exist for IP obfuscation. Other methods deobfuscate IP protected with either combinational~\cite{azar2019smt, chakraborty2021sail} or sequential~\cite{fyrbiak2018difficulty} methods. Sequential obfuscation involves inserting or modifying a control Finite State Machine (FSM). Sequential deobfuscation techniques typically require some level of reverse engineering, which involves identifying control logic and recovering the design's high level function. Fortunately, the same logic recovery that can be leveraged to deobfuscate protected designs can be used to verify a design's logic.  % The control logic is typically a small part of the design, to reduce the overhead incurred by using protection.

% Recovering the high level function requires control logic analysis. More control logic exponentially increases the difficulty and the time required to analyze. The more logic there is in the design, the higher the chance is that it will be infeasible to recover the design's functionality. There are two ways to improve high level function recovery: improve the accuracy of the logic identification, or improve the capabilities of analyzing potential control logic. State-of-the-art control logic identification is somewhat inaccurate in that such methods list non-control logic flip-flops as control~\cite{brunner2019improving}. A 99\% classification accuracy of a design with millions of data flip-flops would be expected to erroneously report tens of thousands of data flip-flops as logic. 

Retrieval of high-level function requires control logic analysis. The complexity of the analysis of such control logic grows exponentially with respect to the retrieved flip-flops, quickly making functional recovery unfeasible. Two avenues for refining high-level control logic recovery exist: bolstering logic identification precision or enhancing potential control logic analysis. Cutting-edge control logic identification exhibits inaccuracies, by classifying non-control logic as control~\cite{brunner2019improving}. Achieving even a 99\% classification accuracy in designs comprising millions of data flip-flops results in reporting tens of thousands of data flip-flops as control.

In the original Register-Transfer Level (RTL), the control FSM states and transitions are explicitly specified. Standard encoding practices yield logarithmic bit usage relative to FSM's state space. Many standard control designs, as represented by those analyzed in this paper, utilize between 3 to 5 flip-flops for state representation, %which would infer that the tens of thousands of  flip-flops would be several orders of magnitudes larger than the number of flip-flops that are actually control even when accounting for the potential for multiple possible logical components within the same design. 
In these same designs the control flip-flops are outnumbered by a factor of tens of thousands of non-control flip-flops. Without methods to further reduce the pool of possible logic and exponential growth of some recovery methods, the storage and analysis of the resulting state machine would be intractable.

Flip-flop interaction analysis methods exist that could reduce the number of flip-flops within groups~\cite{shi2010highly}. Reduced group sizes can enable attackers to analyze pieces of the design for potential control behavior. However, the number of pieces might be large and running recovery tools even once can take some time. To help improve the high level function recovery time of an unknown design the authors of this paper present a workflow and new tool.

The key contributions of this paper are as follows:
\begin{itemize}
    \item Propose using gate-level netlist pruning of designs prior to extracting logic.

    \item Introduce a novel FSM topology extractor, REFSM-SAT, based on solving SAT problems.
    
    \item Demonstrate the capabilities of the proposed FSM extraction method compared to prior methods.

    \item Discuss how topology based FSM extractors can be used to weaken hardware obfuscation.

    % \item We propose a method to cut out an FSM module from a gate level netlist, which can allow for reinsertion after modifications to the FSM.
    
    % \item We propose a modified FSM watermarking scheme that needs less high-level information than prior works.
    
    % \item We propose a method to automatically re-encode an FSM to avoid a class of low level fault types.
    
    % \cut{\item We propose a simplified solution for mitigating a class of Hardware Trojans using our proposed FSM module extractor}
    
    % \item We demonstrate how each of the proposed methods works on real world designs.
\end{itemize}

% In Section~\ref{sec:Related works} we discuss in detail the various problems affecting current solutions to many hardware security challenges. Further in that section we argue that each of these problems can be fixed using a similar design automation tool that helps streamline the processes for IP protection. Several other hardware security problems can leverage the RECUT framework, but are not presented in this work due to page restrictions. Section~\ref{sec:Solution} discusses our proposed technique for addressing the problems laid out in Section~\ref{sec:Related works}. Section~\ref{sec:Methods} describes how we can use the solution for the discussed problems. An analysis of the methods are presented in Section~\ref{sec:Results}. Lastly, Section~\ref{sec:Conclusion} summarizes the findings of this paper.

\section{Related Works}
\label{sec:related}

% Obfuscation Types
The literature categorizes gate-level netlist obfuscation into two primary categories.
%   Combinational
The first category includes methods in which an encrypted design is unlocked with a fixed key. A key is applied by setting design inputs to the key value. Incorrect key applications can flip the value of internal signals~\cite{yasin2015improving}, change the behavior of particular gates~\cite{kamali2018lut}%cocchi2014circuit}~\footnote{Gate camouflaging can be thought of as using the key to choose the correct Boolean function (AND, OR, XOR) of gates at critical points in the design.}\travis{Gate camouflage is not a good method to metnion as it does not prevent overproduction}
, or modify the topology of the combinational circuit~\cite{shamsi2017cyclic}.
%   Sequential
The second category includes methods in which an encrypted design is unlocked with a dynamic key. Some applications of such a design require cycling a set of input pins through a series of key values at startup~\cite{chakraborty2009harpoon}. Another method with dynamic keys involves a fabricated FSM periodically entering a subset of locking states, which requires a secondary FSM executing synchronously to input an unlocking sequence to reacquire function~\cite{hu2020sanscrypt}. Incorrect key applications can provide a denial of service from the correct high-level netlist functions.
%       Emphasize Sanscrypt

% Attack Model
% Combination deobfuscation
%   "Fast" SAT attacks
Recovering the function of an IC protected with a fixed key scheme relies on an unlocked functioning black box version (oracle) of the IC~\cite{subramanyan2015evaluating}. Due to the intended function of the fixed key on the IC, different keys can create different netlist behavior. The same inputs applied to the IC with different keys can create different outputs. These inputs are referred to as Discriminating \addressed{\jim{the literature uses both distinguishing and discriminating, we should be internally consistent}\travis{I assume we will use discriminating?}\Jim{yes}} Input Patterns (DIPs). Using methods discussed in more detail in section \ref{sec:Solution}, DIPs are identified using efficient SAT solvers. Confirmation of the correct behavior is established using the black box IC, which enables identification of a key that is equivalent to, if not the same as, the original intended key.

% Sequential obfuscation/deobfusaction
%   "Fast" SAT attack ignore conditions, but the topology can be useful to determine if state machine is likely non-data
% There is notably less research on recovering designs protected using dynamic keys. Recovery methods will depend on how the dynamic keys are leveraged. If the keys are applied only at startup to unlock the IC~\cite{desai2013interlocking}, then a circuit can be ``unrolled'' and the key sequence can be treated as a longer fixed key~\cite{hu2021fun}. The previously mentioned fixed key, oracle guided method can then be used to discover the longer fixed key. Alternatively, the dynamic key could prevent access to the golden model's state machine by trapping the IC's state in some obfuscated Finite State Machine (FSM)~\cite{chakraborty2009harpoon}. If the encryption method is known and the signals that comprise the FSM are known, then FSM recovery methods~\cite{brunner2019improving, fyrbiak2018difficulty} can help recover the unlocking sequence through examination of the transitions. In some cases, the unlock sequence must cause the design's FSM to transition out of a set of encrypted states to its functional states; this sequence may even be randomized based on a shared secret seed. In this case, an external circuit capable of reproducing the sequence is required~\cite{hu2020sanscrypt}. FSM recovery methods are capable of assisting in identifying the dynamic key through analysis of the state space.

Limited research addresses recovery of dynamic-key-protected designs. Recovery methods hinge on dynamic key utilization. Keys soley unlocked during startup~\cite{desai2013interlocking} can be treated as an extended fixed key~\cite{hu2021fun}. For ICs obfuscated by preventing golden state access~\cite{chakraborty2009harpoon}, FSM recovery can be used if encryption methods and FSM signal composition are known~\cite{brunner2019improving, fyrbiak2018difficulty}. When the FSM transitions between encrypted and function states, an external circuit capable of emulating the state of the protection is needed~\cite{hu2020sanscrypt}. Existing FSM recovery methods are capable of discerning correct logical components through analysis of the state space. %TODO CITE OUR PAPER

%   DPLL to extract conditions

% refsm3
Several methods for enumerating FSM transition graphs have been described in the literature. A state transition can be discovered by constraining the inputs of the FSM until the inputs of the state registers themselves are fixed; careful selection of which inputs are constrained can eliminate the need to search the entire input space~\cite{meade2016refsm}. The state register input fanins can be converted into binary decision diagrams (BDD's) and then brute forced, breaking out when assignments become forced; however, this does not scale to larger FSM's or FSM-like structures such as counters~\cite{highway_to_hal}. AVFSM~\cite{AVFSM} builds a state transition graph by applying inputs to an enhanced version of an FSM implementation; however, it requires the FSM synthesis report as well as the implementation.

\section{Motivation}
\label{sec:motivation}

%%%%%%%%%%%%%%%%% TOPOLOGY %%%%%%%%%%%%%%%%% 

%% Explain topology extraction.
For the purposes of this paper an FSM will be defined as a Moore machine -- an abstract discrete structure which stores an internal state that is modified in each iteration. Formally, an FSM is defined by the following pieces of information:
\begin{enumerate}
    \item A set of possible states the machine can obtain ($S$)
    \item A set of possible inputs the machine can process ($I$)
    \item A state that the machine starts in, which is a member of the set of possible states ($S_0 \in S$)
    \item A transition function which determines the state the machine will achieve in the next iteration based on the state of the machine in the current iteration and the input that is given to the machine in the current iteration ($Tr : S\times I \rightarrow S$)
\end{enumerate}
% Note that this definition of an FSM differs from the discrete mathematics definition; our definition does not include a set of final ``accept''/``reject'' states. In hardware, an FSM typically continues running for as long as power is applied. 
This paper omits the accept state set that the mathematical definition of an FSM typically includes.
The topology of the FSM for the purpose of this paper is defined to be a modification of the transition function, where the input to the resulting function is the current state of the machine and the output is the set of all states that \textit{can} be reached given the proper input $(To : S \rightarrow \mathcal{P}(S))$. %, moreover, 

%$$\forall s,t \in S(\exists i \in I~Tr(s,i) = t \rightarrow t \in To(s)) $$

%% Time Improvement
The topology of an FSM requires less storage than the full transition function, which can be useful if many FSMs are extracted. Discovering topology only requires finding one input word per transition; complete condition extraction demands thorough input space exploration. Additionally, in transition extraction preservation of expression complexity is required to ensure comprehensive condition recovery between each pair of states. %To extract the conditions of a transition, some expression reductions would not be allowed, to ensure that the full set of conditions between two states can be found.

Extracting each transition created in the FSM product is computationally expensive, and determining the transition condition for each transition in the FSM product multiplies the recovery by a factor that tends to be exponential with respect to the number of inputs. Topology extraction of the FSM can be useful when determining the potential for a group of signals to comprise an FSM. 
Combining independent signals into a single FSM tends to form dense transition graphs. Transition graph heuristics like those proposed in REFSM~\cite{NETA} can be used to help split larger words. Knowledge of the defense and how it can modify the topology of the FSM helps determine whether an extracted FSM has the hallmarks of a protection FSM. Topology extraction can additionally provide a way to quickly find potential extraneously added or undefined transitions (i.e., transitions which happen due to accidents of implementation which are not specified in the original design) if the high level state graph is known.

This paper uses methods involving blocking clauses inspired by other SAT solver attacks on combinational obfuscation to extract FSM topologies. The discriminating input algorithm is modified to find different achievable states as opposed to different outputs.

%%%%%%%%%%%%%%%%% REDUCED NETLIST ANALYSIS %%%%%%%%%%%%%%%%% 
Lastly, to reduce the problem size given to the SAT solver, a method for stripping away non-affecting gates should be employed. Previous works proposed methods that operate on a reduced part of gate-level netlist, but based on resulting baseline performance analysis, such methods at least store if not evaluate unnecessary gates. Keeping additional gates in the program's memory can negatively affect cache performance, as well as slow down solvers which do not have data on which part of the netlist is of the deobfuscation method's interest. Motivated by the possibility of improving runtime, this paper also presents a comparison of using a reduced netlist as inputs for the presented FSM and FSM topology extraction methods to the same methods given the raw gate-level netlist.

\section {Proposed Solution}
\label{sec:Solution}

\subsection{RECUT}

\addressed{\travis{For the following sections try to avoid the word we if possible.}}
Cutting the cells that compose the FSM from a design starts with identifying the FSM's state registers\addressed{~\travis{what are these stages?}}. The problem of finding the state registers is well-studied, and is done using existing tools~\cite{meade2016, geistreflic}. \cut{\jim{we aren't running REFSM yet If the original design is available, it can be used to cross-check the states and transitions found by the tools.}}

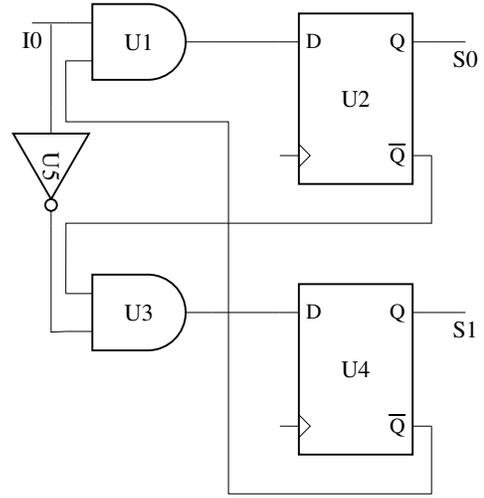
\begin{figure}
    \centering
    \scalebox{0.90}{
    \begin{circuitikz}
    \draw (0,0) node[flipflop D](S0){U2} (S0.bup);
    \draw (S0.pin 1) -- ++(-1,0) node[and port,anchor=out](AND1){U1};
    
    \draw (0,-4) node[flipflop D](S1){U4} (S1.bup);
    \draw (S1.pin 1) -- ++(-1,0) node[and port,anchor=out](AND2){U3};
    
    \draw (S0.pin 4) -- ++(0,-1) -- ++(-5.4, 0) -- (AND2.in 1);
    \draw (S1.pin 4) -- ++(0,-1) -- ++(-3.0, 0) -- ++(0, 5.5) -- ++(-2.4,0) -- (AND1.in 2);
    
    \draw (S0.pin 6) -- ++ (0.5, 0) node[below]{S0};
    \draw (S1.pin 6) -- ++ (0.5, 0) node[below]{S1};
    \draw (AND1.in 1) -- ++ (-0.5, 0) node[below]{I0};
    
    \draw (-4.5,-1) node[not port, rotate=270](NOT1){U5};
    \draw (-4.5, 1.1) -- (NOT1.in);
    \draw (NOT1.out) -- (-4.5, -3.45) -- (AND2.in 2);
    \end{circuitikz}
    }
    \caption{Example FSM as a gate level netlist.}\vspace{0.1in}
    \label{fig:cut1}
\end{figure}

Some works define a gate-level netlist FSM implementation as the registers implementing the current state along with any combinational logic in a path leading from a state register output to a state register input. The inputs to an FSM are defined as the minimal set of signals which, when combined with the signals representing the FSM's current state, totally determine the FSM's next state. However, using this definition to extract a netlist cut can lead to a cut which is underconstrained. In figure \ref{fig:cut1}, gates $U1$ and $U3$ meet the criteria, but inverter $U5$ does not. $U5$ constrains the outputs of $U1$ and $U3$ such that they cannot both be simultaneously true, and omitting it will lead to the FSM, in isolation, implementing transitions which are not possible with the FSM \textit{in situ}. An underconstrained result is undesirable because it can lead to misunderstanding of the FSM's intent.%, and because state enumeration will take longer as there are more transitions to explore.

Under the assumption that the next state of the FSM only depends on the current state registers (i.e. non-state registers can take on arbitrary values), the cut criteria includes the state registers and any combinational logic which drives their inputs. In practice, this criteria means including the union of the state registers' fan-in cones, terminating in the outputs of other registers or primary inputs. This definition creates cuts which implement the same transitions as the FSM in the full design. In some cases, as shown by experimentation in Section \ref{sec:results}, extracting the transitions from the cut takes far less time than from the full design.

Algorithm~\ref{alg:maxcut} shows the method for computing the cut: a breadth-first search (BFS) of the fan-in cones of the state registers, similar to methods described in \cite{mcelvain}. As BFS is linear time ($O(V+E)$) taking a cut is extremely fast even on large netlists.

\addressed{~\travis{try to avoid the word ``this''.} case, the algorithm returns the set of cells which contains all identified gates in the cut.}

\begin{algorithm}
  \caption{Extraction of an FSM Cut from a Gate-Level Netlist}
  \begin{algorithmic} 
    \small
    \Function{FSMcut}{$FSM$}
    \State {$Work \leftarrow$ Queue}
    \State {$Cut \leftarrow$ Set}
    \State {Enqueue all FSM DFF's into $Work$}
    \While {$Work$ is not empty}
      \State {Pull $Curr$ from $Work$}
      \If {$Curr \notin Cut$}
        \State {Add $Curr$ to $Cut$}
        \If {$Curr$ is not a primary input or register output}
            \State {Enqueue $Curr$'s fanin signals to $Work$}
        \EndIf
      \EndIf
    \EndWhile
    \State {Return $Cut$}
    \EndFunction
  \end{algorithmic}
  \label{alg:maxcut}
\end{algorithm}

\subsection{FSM Topology Enumeration}

As discussed in Section~\ref{sec:related}, SAT problems have been used to break combinational and sequential logic schemes by discovering a key needed to unlock a circuit. In combinational locking, DIPs can be used to quickly narrow down the set of possible keys to a locked circuit; if two key input patterns produce different outputs, then only one can be the correct key. By iteratively adding SAT clauses to rule out incorrect keys, the problem can be narrowed down to the correct key~\cite{subramanyan2015evaluating}. If the locking mechanism contains state elements (flip-flops), then key inputs over multiple clock cycles must be solved. One way to transform such a sequential circuit into a SAT problem is to replicate the design once for every clock cycle; the register inputs of the copy of the design representing clock cycle $n$ become the register outputs in the copy of the design representing cycle $n+1$. If there are $n$ clock cycles and $k$ key bits, the solution becomes a bit vector containing $nk$ bits, representing the key inputs over each cycle. This is known as unrolling the circuit~\cite{KC2}.

\begin{figure}
\centering
\scalebox{1.0}{\begin{tikzpicture}

%\draw [help lines] (0,0) grid (8, 6);
%\draw (0, 0) rectangle (7.75,5.25); % outer rect

% combo logic cloud
%
\node (comblogic) [cloud, draw, fill=white, cloud puffs=10,cloud puff arc=120, aspect=1.5, xscale=2.2, yscale=2.0, inner ysep=1em] at (2.5, 2.5) {}; % combo logic
\node at (2.5, 2.5) {Combinational Logic};

\node (eq1) at (4,0) {Current state = 101: clauses $Q_0$, $\lnot Q_1$, $Q_2$};
\node (eq2) at (4, -0.30) {Next state $\ne$ 001: clause $D_0 \lor D_1 \lor \lnot D_2$};

% split registers
%
\node (d0) [draw, label={$R_0$}, minimum width=1.0cm, minimum height=1.0cm, fill=white] at (5.5,1) {$D_0$};
\node (q0) [draw, minimum width=1.0cm, minimum height=1.0cm, fill=white] at (6.5,1) {$Q_0$};
\node (d1) [draw, label={$R_1$}, minimum width=1.0cm, minimum height=1.0cm, fill=white] at (5.5,2.5) {$D_1$};
\node (q1) [draw, minimum width=1.0cm, minimum height=1.0cm, fill=white] at (6.5,2.5) {$Q_1$};
\node (d2) [draw, label={$R_2$}, minimum width=1.0cm, minimum height=1.0cm, fill=white] at (5.5,4) {$D_2$};
\node (q2) [draw, minimum width=1.0cm, minimum height=1.0cm, fill=white] at (6.5,4) {$Q_2$};

% output wires
%
\node (o0) at (8.25, 1) {$O_0$};
\node (o1) at (8.25, 2.5) {$O_1$};
\node (o2) at (8.25, 4) {$O_2$};

% input wires
%
\node (in) at (0.25, 2.5) {$I_n$};

% comb logic to D side
%
\draw [->] (comblogic) -- (d0);
\draw [->] (comblogic) -- (d1);
\draw [->] (comblogic) -- (d2);
\draw [->] (in) -- (comblogic);

% Q side to output
%
\draw (q0) -- (o0);
\draw (q1) -- (o1);
\draw (q2) -- (o2);

% feedback loops up
%
\draw (7.25,4) -- (7.25, 5.25);
\draw[fill, black] (7.25, 4) circle[radius=0.05cm];
\draw (7.5,2.5) -- (7.5, 5.5);
\draw[fill, black] (7.5, 2.5) circle[radius=0.05cm];
\draw (7.75,1) -- (7.75, 5.75);
\draw[fill, black] (7.75, 1) circle[radius=0.05cm];

% feedback loops - right to left
%
\draw (7.25, 5.25) -- (2.75, 5.25); 
\draw (7.5, 5.5)   -- (2.5, 5.5); 
\draw (7.75, 5.75) -- (2.25, 5.75); 

% feedback loops - down
%
\draw [->] (2.75, 5.25) -- (2.75, 3.6);
\draw [->] (2.5, 5.5) -- (comblogic);
\draw [->] (2.25, 5.75) -- (2.25, 3.6);

\end{tikzpicture}}
\caption{SAT Problem Setup. The clauses $Q_0$, ${\lnot}Q_1$, and $Q_2$ force a solution that has the $Q$ signals set to $101$. The clause $D_0 \lor D_1 \lor \lnot D_2$ forces a solution where at least one bit of the $D$ signals differs from $001$.\addressed{~\travis{Can this caption explain the current/next state in more detail?}}}
\label{fig:miter}
\end{figure}

REFSM-SAT enumerates transitions by solving a slightly different SAT problem. Figure \ref{fig:miter} shows a typical FSM with inputs $I_n$ and outputs $O_n$. In the combinational logic section, there are signals and combinational gates. Each signal is represented by a SAT variable, and each gate is represented as a clause constraining the gate's inputs and output based on the gate's function. Each register $R_i$ is represented by two variables: $D_i$ representing the register's input value, and $Q_i$ representing the register's output. On each clock cycle in the hardware implementation, $Q_i$ will be set to $D_i$, and $D_i$ will change based on the combinational logic. Therefore the $Q_i$ variables contain the current state and the $D_i$ variables the next state. To enumerate possible next states for a current state $S$, clauses constraining $Q$ to be $S$ are added to the problem; for example, if $S = 101$, the clauses $Q_0$, $\lnot Q_1$, and $Q_2$ are added. If there is a solution to the SAT problem, then the solver software can be queried for the satisfying values of the $D_i$ variables, which will give a next state $N$; if there is no solution, then there are no more next states to find from the current state $S$. In order to find another possible next state, this state must be excluded by adding another clause to constrain $D_i \ne N$. For example, if $N = 001$, the clause $D_0 \lor D_1 \lor \lnot D_2$ is added. By iterating in this manner until there is no solution, all possible next states of $S$ can be found.

With a method for determining the set of next states from any current state, it is trivial to enumerate all transitions. The states and transitions form a directed graph (the FSM topology) with the states as nodes and the transitions as edges. The start state is the value of each $Q_i$ variable when the reset signal is asserted. A BFS can then be used to enumerate the entire graph, as shown in Algorithm~\ref{alg:satfinder}. 
The function $SetEqual$ adds blocking clauses to the SAT problem such that the variables (signals) have the given values (i.e., clauses for each signal or its complement if the signal value must be one or zero, respectively; $SetNotEqual$ adds a blocking clause that the variables (signals) must not all match the given value (i.e., a clause where each term is the negation of the given value).

It is important to note that the original REFSM algorithm cannot stop searching for transitions from a state until all possible input conditions have been enumerated. In the worst case, the number of input combinations is exponential in the number of inputs; precisely, it is $2^I$ where $I$ is the number of inputs. Intelligent culling of the input space to account for don't care signals can reduce the space to tractable levels. For example, if an input to an AND gate is set to zero, then the other gate inputs can be treated as don't care. However, complete coverage (including those from don't cares) of the input space is still required. The REFSM algorithm must enumerate all conditions and report each discovered state only once. This allows REFSM to list complete coverage of the input conditions leading to each transition; however, the conditions are not always necessary for all analyses. For example, when looking for obfuscation or locking, only the structure of the topology is needed.~\cite{NETA}

\cut{
\addressed{\travis{This paragraph feels like it should occur later in this section.}}
In order to determine the correctness of cuts created with RECUT, and to measure performance of FSM enumeration on both the original netlist and the cut, REFSM~\cite{meade2016refsm} was run on both the original netlist and the cut. \addressed{We were also curious~\travis{too informal. Maybe refer to the motivation.} if the REFSM algorithm itself could be improved.} Per the original work, REFSM searches the transition space for each state by searching through the fan-in space of each state register and setting wires to true or false states. Upon setting a new wire, a constraint solver is used to check if the input values to all state registers are fully determined; if so, then the state register inputs \addressed{~\travis{vague}} are a possible next state. As discussed in section \ref{sec:motivation}, while \addressed{this algorithm~\travis{reference?}}REFSM allows enumeration of the conditions of each transition as well as the states themselves, a more general SAT solver with better heuristics might outperform REFSM if transition conditions were not required.}

\begin{algorithm}
  \caption{Enumeration of FSM via SAT solver}
  \begin{algorithmic} 
    \small
    \Function{FSMSolve}{$FSM$, $ResetState$}
    \State {$StateWord \leftarrow FSM.Registers$}
    \State {$D \leftarrow StateWord.InputSignals$}
    \State {$Q \leftarrow StateWord.OutputSignals$}
    \State {$Work \leftarrow$ Queue}
    \State {$SeenStates \leftarrow$ Set}
    \State {$Transitions \leftarrow$ Set}
    \State {Enqueue($Work, ResetState$)}
    \While {$Work$ is not empty}
      \State {$Curr \leftarrow $Dequeue($Work$)}
      \If {$Curr \notin SeenStates$}
        \State {Add($SeenStates, Curr$)}
        \State {$SATProblem \leftarrow$ ComboLogicToSATClauses($FSM$)}
        \State {SetEqual($SATProblem, Q, Curr$)}
        \While {$Solution \leftarrow SATProblem.Solve()$}
            \State {$Next \leftarrow Solution(D)$}
            \State {Add($Transitions, Curr \rightarrow Next)$}
            \State {SetNotEqual($SATProblem, D, Next$)}
            \State {Enqueue($Work, Next$)}
        \EndWhile
      \EndIf
    \EndWhile
    \State {Return $Transitions$}
    \EndFunction
  \end{algorithmic}
  \label{alg:satfinder}
\end{algorithm}

\section{Experimental Results}
    \label{sec:results}
    
\subsection{Benchmarks and Experimentation Setup}

\travis{Include details that specify REFSM 2 is used for benchmarking against the new version of REFSM.}

Table~\ref{tab:results-netlists} shows the netlists used for performance measurement. The netlists cover a spectrum of designs and topologies. There are two RISC-V cores: ibex, a 32-bit core designed for embedded control applications~\cite{ibex}, and cv32e40s, another 32-bit core designed for security applications~\cite{cv32e40s}. Two netlists from ISCAS~\cite{iscas-circuits} are included: s13207 and s38417, which are scans of real designs. The remaining netlists are from OpenCores~\cite{opencores}. The RSA design is an encryption core. The two ETHMAC netlists are the receive and transmit side of an Ethernet controller, PCI is a PCI bus controller, and UART is an RS-232 controller. The MC8051 is a full microcontroller.

RECUT and REFSM-SAT are implemented in C++. All experiments were run on a 3.30GHz Intel Core i7-5820K desktop with 64GB of memory. REFSM-SAT is based on the Minisat 2.2 SAT solver~\cite{minisat}. Table~\ref{tab:results-perf} shows the performance results. For each netlist, a state word was extracted using common tools. REDPEN and RELIC from the NetA toolset~\cite{NETA} were used to find likely state words for each netlist. RECUT was run to extract the FSM. REFSM 2 and REFSM-SAT were then run on both the original netlist and the cut. REFSM 2, which improves performance by pruning the FSM's input signals using dominator trees, is the most current version of classic REFSM.
%\subsection{Example use case}
\addressed{\jim{We spend very little time talking about this example, and have little room to expound on it more -- do we need it?}~\travis{Probably not. Let's first see how we are doing on room.}~\travis{Upon further thinking, I feel it is important to keep, since the paper mentions validating logic in the motivation. We should probably move the material on bad UART into the Experiments section}}
%%Figure~\ref{fig:eviluart} shows one use case where conditions are not required; investigating the FSM for a simple UART transmitter. The transitions represented by solid black lines are expected due to the original design. In this case, a malicious actor has inserted an extra transition to corrupt the output of the UART in some cases; that transition is shown as a red, dashed line. This transition will cause the UART to reset in the middle of a byte, causing the receiver to lose sync. \travis{I think we should remove this sentence since we are not using hardware Trojans} % This is enough to set up a side channel information leak.
%%Transition conditions are not needed to find this extra transition. 
%%\input{figures/uart}
\begin{table*}
  \begin{tabular}{|l||c|c|c|c||c|c|c|c|c|}
    \hline
    & \multicolumn{4}{c||}{Full Netlist} & \multicolumn{4}{c|}{RECUT Sub-Netlist}&\\
    Netlist & Inputs & Regs & Gates & Cells & Inputs & Regs & Gates & Cells & ACPT \\
    \hline
   % aes\_host13\_pr & 393 & 390 & 1576 & 1448 & 3 & 4 & 24 & 23 & 1.00 \\
    cv32e40s & 283 & 2352 & 43100 & 20310 & 311 & 0 & 844 & 1117 &  \\
    ethmac-rxstate & 12 & 6 & 34 & 28 & 8 & 6 & 34 & 28 & 1.21 \\
    ethmac-txstate & 59 & 12 & 143 & 99 & 51 & 11 & 143 & 98 & $1.04{\cdot}10^4$ \\   % 10392.40 \\
    ibex & 183 & 1932 & 28195 & 15335 & 1222 & 1206 & 10422 & 5092 & $5.59{\cdot}10^6$\\
    mc8051 & 92 & 578 & 10408 & 6590 & 39 & 3 & 857 & 689 & $1.43{\cdot}10^4$ \\ %%  14263.89 \\
    pci & 57 & 33 & 238 & 169 & 28 & 3 & 47 & 40 & 400.24 \\
%    pdp8-fsm & 26 & 5 & 101 & 75 & 26 & 5 & 101 & 75 & 1.76 \\
    rsa & 99 & 555 & 3967 & 2139 & 99 & 3 & 92 & 92 & $1.69{\cdot}10^2$ \\ %% 168.50 \\
    s13207 & 63 & 638 & 7951 & -- & 4 & 13 & 193 & -- & 1.29 \\
%    s1494 & 9 & 6 & 647 & -- & 9 & 6 & 320 & -- & 1.00 \\
%    s344 & 10 & 15 & 160 & -- & 13 & 8 & 116 & -- & 1.89 \\
    s38417 & 29 & 1636 & 22179 & -- & 3 & 7 & 115 & -- & 1.00 \\
%    s444 & 4 & 21 & 181 & -- & 11 & 4 & 44 & -- & 9.10 \\
%    s510 & 20 & 6 & 211 & -- & 20 & 6 & 173 & -- & 1.01 \\
%    s641 & 36 & 19 & 379 & -- & 16 & 15 & 259 & -- & 4.79 \\
%    s713 & 36 & 19 & 393 & -- & 16 & 15 & 272 & -- & 4.67 \\
%    s9234 & 37 & 211 & 5597 & -- & 4 & 5 & 33 & -- & 3.08 \\
%    s953 & 17 & 29 & 395 & -- & 15 & 6 & 208 & -- & 4.69 \\
    uart\_1 & 12 & 59 & 181 & 168 & 12 & 3 & 27 & 24 & 4.18 \\
    uart\_2 & 12 & 59 & 181 & 168 & 11 & 3 & 21 & 21 & 2.14 \\
    \hline
  \end{tabular}
      \vspace{0.1in}
  \centering 
  \caption{Netlist data set. ACPT is the average number of conditions per transition.}
  \label{tab:results-netlists}
\end{table*}

\begin{table*}
  \begin{tabular}{|l|c|c|c|c|c|c|c|}
    \hline
    & & \multicolumn{3}{|c}{Full Netlist} & \multicolumn{2}{|c|}{RECUT Sub-Netlist} & Overall \\
    Netlist & RECUT & REFSM & REFSM-SAT & $\Delta$ & REFSM & REFSM-SAT & $\Delta$ \\
    \hline
 % aes\_host13\_pr & 0.01 & 0.01 & 0.05 & 0.04 & 0.01 & 0.01 & 0.00\\
  cv32e40s & 0.13 & N/A & 0.79 & N/A & N/A & 0.05 & N/A \\ 
  ethmac-rxstate & 0.01 & 0.01 & 0.01 & 0.00 &  0.01 & 0.01 & 0.00 \\
  ethmac-txstate & 0.01 & 18.08 & 0.11 & -17.97 & 17.99 & 0.11 & -17.97 \\
  ibex & 0.11 & 2214.52 & 0.31 & -2214.21 & 594.59 & 0.13 & -2214.39 \\
  mc8051 & 0.05 & 14.52 & 0.12 & -14.40 & 0.15 & 0.03 & -14.94 \\
  pci & 0.01 & 0.02 & 0.01 & -0.01 & 0.01 & 0.01 & -0.01 \\
 % pdp8-fsm & 0.01 & 0.01 & 0.02 & 0.01 & 0.01 & 0.01 & 0.00 \\
  rsa & 0.02 & 0.07 & 0.08 & 0.01 & 0.01 & 0.01 & -0.06 \\
  s13207 & 0.03 & 0.09 & 13.80 & 13.71 & 0.05 & 0.44 & 0.35 \\
%  s1494 & 0.01 & 0.01 & 0.01 & 0.00 & 0.00 & 0.01 & 0.00\\
%  s344 & 0.00 & 0.04 & 0.18 0.14 & 0.04 & 0.14 & 0.10 \\
  s38417 & 0.06 & 0.07 & 1.56 & 1.49 & 0.00 & 0.02 & -0.05 \\
%  s444 & 0.00 & 0.00 & 0.01 & -- & 0.00 & 0.01 & 0.10 \\
%  s510 & 0.00 & 0.00 & 0.02 & -- & 0.00 & 0.02 & -- \\
%  s641 & 0.01 & 0.31 & 1.40 & 1.09  & 0.28 & 1.02 & 0.71 \\
%  s713 & 0.01 & 0.30 & 1.50 & 1.20 & 0.31 & 1.13 & 0.83 \\
%  s9234 & 0.02 & 0.03 & 0.05 & 0.02 & 0.00 & 0.00 & -0.03 \\
%  s953 & 0.01 & 0.01 & 0.02 & 0.02 &  0.00 & 0.01 & 0.00 \\
  uart\_1 & 0.01 & 0.01 & 0.00 & 0.01 & 0.01 & 0.01 & 0.00 \\
  uart\_2 & 0.01 & 0.01 & 0.00 & 0.01 & 0.01 & 0.01 & 0.00 \\

    \hline
  \end{tabular}
      \vspace{0.1in}
\centering
    \caption{Runtimes of RECUT, REFSM, and REFSM-SAT on original and FSM cut netlists. All times are in seconds. Full Netlist $\Delta$ is the performance change of REFSM-SAT versus REFSM on the full netlist. Overall $\Delta$ is the performance change of REFSM-SAT on the RECUT netlist versus REFSM on the full netlist.}
  \label{tab:results-perf}

\end{table*}

\subsection{REFSM-SAT Results}
On all designs where REFSM completed, the extracted topologies were identical between REFSM and REFSM-SAT. REFSM and REFSM-SAT show little performance difference on small netlists. On two netlists with a high number of conditions per transition, ETHMAC-TXSTATE and MC8051, REFSM-SAT shows significant benefit, running 99\% faster. Running REFSM with conditions enabled shows that these two netlists have a much higher average number of conditions per transition (where each condition is one set of signal assignments which will cause the transition) than the other test cases, with over 10,000 conditions per transition. On the two RISC-V netlists, ibex and cv32e40s, the difference is even more profound. On the ibex design, REFSM takes 36 minutes to enumerate the transitions and REFSM-SAT only 310 milliseconds. On the cv32e40s design, REFSM did not complete in 4 days. Due to the state machine in the ibex CPU depending on a large amount of state from the entire design, the average number of conditions per transition is $5.59{\cdot}10^6$; the transition statistic is not available for cv32e40s as computing the statistic requires the REFSM results. These designs exemplify the pathological case that REFSM-SAT was designed to combat. In general, if a design is expected to have many conditions per transition, or very complex fan-in cone logic (such as is often found in FSM's in obfuscation schemes~\cite{hu2020sanscrypt}), then REFSM-SAT is a better choice than REFSM.

\subsection{RECUT Results}
The RECUT algorithm extracts a cut from the netlist which should, given the fact that REFSM and REFSM-SAT only produce results based on the current and next clock cycles, produce identical transition graphs to the full netlist for the state word used to generate the cut. Furthermore, using RECUT as a preprocessing step may provide performance benefits, as there is less data to deal with and the solvers may evaluate parts of the full netlist which do not affect the transition space, as shown in table~\ref{tab:results-perf}. In general, evaluating the netlist cut versus the full netlist takes a comparable amount of time for smaller netlists. For the two medium size netlists (MC8051 and s38417) the speedup is significant. In both cases, REFSM is 99\% faster on the cut than the full netlist. REFSM-SAT is 75\% faster on the MC8051 cut, and 99\% faster on the s38417 cut. 

On the largest netlists, ibex and cv32e40s, the cost of running RECUT as a preprocessing step is still negligible (under 150 ms). Enumerating the transitions on the RECUT output of ibex is significantly faster than on the unmodified netlist with both tools: REFSM is 73\% percent faster, and REFSM-SAT is 58\% faster; on the cv32e40s design, REFSM did not complete in multiple days even with RECUT as a preprocessing step. On the RECUT output of ibex, REFSM-SAT is 99\% faster than classic REFSM. On larger netlists, there is processing overhead in loading and storing the entire netlist. In REFSM, simulation may reach into areas of the netlist which do not change the transition outcome; in REFSM-SAT, the larger netlist leads to a larger SAT problem. In both cases, a smaller data set leads to better performance with negligible preprocessing time.

\section{Conclusion}

% Not all the same XXX and YYY
% Talk about related works `` Past methods of XXX and YYY were shown. ''
% Talk about motivation `` The method of XXX was justified for reason of YYY ''
% Talk about algorithm `` Methods for XXX and YYY that used were presented ''
% Talk about result. `` It was shown that REFSM had XXX performance, and REFSM-SAT had YYY performance, and RECUT had ZZZ peformance''

% \jim{should this be here? we've shifted the paper to be about performance, and this is the only section that talks about removal and reinsertion}
% \travis{this was moved and needs polish}
% If the use case is to transform the cut and reinsert it into the design, the inverse cut can easily be found by taking the complement of the cut gates from the full netlist gate set.
% If the original netlist is specified at the gate level, then algorithm~\ref{alg:maxcut}\addressed{~\travis{which algorithm? reference it.}} will work as shown. If the netlist is specified in terms of a technology library, then the cut should be taken to respect cell boundaries in order to not require changing formats, and Algorithm~\ref{alg:maxcut} is extended to return all cells which contain any gate in the cut.

In this paper, we address the problem of FSM state enumeration for hardware security. While there are existing tools such as REFSM which can perform state enumeration, we showed that in some cases the REFSM algorithm can be improved\addressed{~\travis{is 18 seconds not well? maybe ``can be improved''}}. We noted that in many cases, state enumeration without transition conditions is a sufficient result; as REFSM always computes transition conditions even when they are not needed, it suffers performance overhead on designs with many different transition conditions.

We introduced two new tools: RECUT and REFSM-SAT. RECUT extracts a cut of a netlist which contains only the gates and registers which directly affect the FSM states and transitions. We showed that running RECUT as a preprocessing step to enumeration can significantly improve performance, as the enumeration algorithm searches through a much smaller netlist. REFSM-SAT formulates the enumeration as a SAT problem, which allows the solver to search for state transitions independently of transition conditions. As the REFSM solver enumerates the condition space and provides state transitions as a side effect of that search, REFSM-SAT is significantly faster on designs with many possible condition transitions, such as CPU cores.

\section*{Acknowledgements}
This work is partially supported by the National Science Foundation (NSF-1812071).

%%% References
{
\footnotesize
\printbibliography
}

\end{document}